%% file: main_VM.tex
\documentclass[11pt]{article}

\usepackage{graphicx}
\usepackage{latexsym,amsmath,amsfonts,amscd, amsthm, dsfont}

\usepackage{bm,color}
\usepackage{epsfig,verbatim,epstopdf,graphics}
\usepackage{subfigure}
\usepackage{changebar}
\usepackage{multirow}
\numberwithin{figure}{section}
\usepackage{xcolor}
\usepackage{url}
	\usepackage[ruled,vlined]{algorithm2e}

	\usepackage{yhmath}
	\usepackage{booktabs} 
	\usepackage{tikz}
	\usepackage{verbatim}
	\usetikzlibrary{arrows,backgrounds,snakes,shapes}
	
	\numberwithin{equation}{section}

	\graphicspath{{./}{./figure/}}
	\allowdisplaybreaks
	
	\newtheoremstyle{plainNoItalics}{}{}{\normalfont}{}{\bfseries}{.}{ }{}
	
	\theoremstyle{plain}

	\theoremstyle{plainNoItalics}

	\newcommand{\beq}{\begin{equation}}
		\newcommand{\eeq}{\end{equation}}
	\newcommand{\bit}{\begin{itemize}}
		\newcommand{\eit}{\end{itemize}}
	\newcommand{\be}{\begin{eqnarray}}
		\newcommand{\ee}{\end{eqnarray}}
	\newcommand{\beno}{\begin{eqnarray*}}
		\newcommand{\eeno}{\end{eqnarray*}}

	\newcommand{\Rmnum}[1]{\expandafter\@slowromancap\romannumeral #1@}
	\makeatother
	
	\usepackage{enumerate}

\begin{document}
\baselineskip=1.5pc

\vspace{.5in}

\input{title}

\input{intro}

\input{Background}
\input{algorithm}

\input{numerical}
\input{conclusion}

\bibliographystyle{siam}
\bibliography{Ref_Sa}

\end{document}

%% file: title.tex
\begin{center}
{\bf
Insights from the Field: A Comprehensive Analysis of Industrial Accidents in Plants and Strategies for Enhanced Workplace Safety.
}
\end{center}

\vspace{.2in}
\centerline{
Hasanika Samarasinghe \footnote{
	Department of Engineering, Sony UK Technology Centre, Pencoed Technology Park, Bridgend, United Kindom, CF35 5HZ. E-mail:                                                                                                                                        	svasha91@gmail.com.
} 
		

and
 	
 Shadi Heenatigala\footnote{
	Department of Mathematics, Texas State University, San Marcos, TX, USA,  78666. E-mail:                                                                                                                                                                          
	ydq14@txstate.edu.
} 
}
\bigskip
\noindent
{\bf Abstract.}
The study delves into 425 industrial incidents documented on Kaggle \cite{FindKaggle}, all of which occurred in 12 separate plants in the South American region. By meticulously examining this extensive dataset, we aim to uncover valuable insights into the occurrence of accidents, identify recurring trends, and illuminate underlying causes. The implications of this analysis extend beyond mere statistical observation, offering organizations an opportunity to enhance safety and health management practices. Our findings underscore the importance of addressing specific areas for improvement, empowering organizations to fortify safety measures, mitigate risks, and cultivate a secure working environment. We advocate for strategically applying statistical analysis and data visualization techniques to leverage this wealth of information effectively. This approach facilitates the extraction of meaningful insights and empowers decision-makers to implement targeted improvements, fostering a preventive mindset, and promoting a safety culture within organizations. This research is a crucial resource for organizations committed to transforming data into actionable strategies for accident prevention and creating a safer workplace.
 
\vfill

\newpage

%% file: intro.tex
\section{Introduction}
Workplace health and safety practices play a crucial role in ensuring the well-being of both the organization and its employees. Ineffective policies in this area can adversely affect the firm and its workforce. While safety and health concepts are globally applicable, the specific actions required may vary based on the organization's size, the hazards associated with its operations, its overall culture, products, or services, and the adequacy of existing arrangements.\\ 
The twentieth century witnessed a significant increase in the risks associated with technology, prompting companies to prioritize the reduction of workplace injuries and illnesses. Consequently, safety management has emerged as a prominent discipline in the business world. Extensive research has been conducted on safety management practices in various industries globally, but there is a lack of studies specifically focused on real-world manufacturing plants.\\ 
The Industrial Revolution in the 18th century indeed brought about significant changes in manufacturing processes, including the introduction of machines to replace human labor. This led to the establishment of large-scale factories where workers were closely supervised and tasks were divided among them. \\
As Felton \cite{Felton1986} points out, the rapid growth of the Industrial Revolution was accompanied by an increase in work-related accidents, resulting in injuries and fatalities. It is important to note that the Industrial Revolution marked a pivotal moment in history, where the focus on productivity and efficiency often overshadowed concerns for worker safety. The lack of proper safety regulations and protective measures during this period contributed to the rise in work-related accidents. Over time, however, societies recognized the importance of workplace safety and began implementing regulations and practices to protect workers. This led to the development of occupational health and safety standards that continue to evolve and improve today.\\
According to Booth and Lee \cite{booth995role}, safety management refers to a process driven by top management aimed at reducing risks to the health and safety of workers. The primary objective of safety management is to identify and address the causes of accidents and incidents. This involves actively identifying both observable and latent hazards. Hazard detection is just one aspect of safety management, encompassing a comprehensive system for planning, implementing, and following up on safety operations. Safety management typically involves risk analysis, coordination of safety training, investigation of accidents and near-misses, safety promotion, and assessment of human reliability. Grimaldi and Simmonds \cite{Grimaldi1975Safty} emphasized that a successful safety management system assigns these tasks to all hierarchical levels within the organization. By implementing a robust safety management system, companies can effectively mitigate risks and ensure the well-being of their employees.\\
This paper is organized as follows. In Section 2, we present the literature review to demonstrate the background of the project. In Section 3, we provide a comprehensive overview of Brazil’s diverse industries, workforce, and socioeconomic landscape. Subsection 3.1 focuses on Occupational Safety and Health Practices in Brazil. In Section 4, we analyze perspectives on industrial accidents. Finally, in Section 5, we conclude by summarizing the main contributions of the paper and offering comments on future research directions.

%% file: Background.tex
\section{Literature Review}
The establishment of the International Labour Organization (ILO) in 1919 \cite{International} marked a significant step towards preventing occupational hazards and creating safe work environments. The ILO, as a specialized agency of the United Nations, is responsible for developing policies, promoting programs, and setting work standards for its 187 member countries. Its primary objective is to ensure the well-being and safety of workers worldwide. In addition to the ILO, the World Health Organization (WHO) \cite{World} also plays a role in addressing occupational health and safety. The WHO focuses on various aspects of health, including occupational health, and works towards promoting and protecting the health of individuals globally. At the European level, the European Agency for Safety and Health at Work (EU-OSHA) \cite{EuropeanAgency} takes the lead in addressing occupational safety and health issues among its member states. It works towards ensuring safe and healthy working conditions for European workers. It is important to note that each country has its own set of laws and regulations for occupational health and safety. These laws and rules are implemented to ensure compliance and protect workers within their respective jurisdictions.\\
According to Cox and Tait \cite{Cox1991Reli}, safety management involves making informed decisions to meet accepted safety standards and achieve a state of freedom from emerging risks or harm. Kennedy and Kirwan \cite{Kennedy1998Development} further explain that organizations employ various safety management strategies to monitor and control work activities and methods within the framework of their safety management system. In developing nations, management voluntarily adopts different safety management strategies to ensure the health and safety of their workers. These strategies are implemented to mitigate risks and create a safe working environment.\\
The studies conducted by Cohen et al. \cite{Cohen1975Safty} and Shannon et al. \cite{Shannon1996Workplace} provide valuable insights into the factors that contribute to lower accident rates and improved workplace safety. In the study by Cohen et al. \cite{Cohen1975Safty}, several factors were identified as influential in reducing accidents, including the presence of high-ranking safety officers, the involvement of leaders in safety activities, regular training for new and current employees, and increased interpersonal interaction between managers and employees. Similarly, the study by Shannon et al.\cite{Shannon1996Workplace} focused on manufacturing enterprises and found that organizations with lower rates of lost time injuries had managers who believed in giving the workforce a greater say in the decision-making process and fostered a harmonious management-worker relationship. These findings highlight the importance of proactive safety measures, leadership involvement, ongoing training, and effective communication between managers and employees in creating a safer work environment. Implementing these practices can help organizations reduce accidents and promote a culture of safety.\\
Researchers have conducted investigations into how safety management techniques impact the safety behaviours of individuals within organizations. One study by Smith-Crowe et al. \cite{Smith2003Organizational} found that safety management methods, particularly safety training, played a moderating role in the relationship between safety awareness and overall safety performance. Another study by Neal et al. \cite{Neali2000impact} utilized structural equation procedures to examine the connection between antecedent factors, determinants, and components of safety performance among hospital workers in Australia. The study revealed that safety awareness and safety motivation individually predicted participation in safety management practices. Additionally, it was found that safety management techniques were predictive of both safety awareness and safety motivation. These findings highlight the importance of implementing effective safety management practices to enhance safety performance within organizations.\\
In industrialized countries, government agencies have indeed guided the creation and operation of effective health and safety management systems. This has led to safety management becoming an established practice in many industries. In several cases, these guidelines have been integrated with existing quality management and system standards, creating a comprehensive framework for organizations to ensure both safety and quality in their operations. These standards and guidelines serve as valuable resources for companies seeking to implement robust health and safety management systems.\\
Ensuring effective safety and health management practices is crucial in hazardous areas like chemical or process industries. According to the study mentioned by Hurst et al. \cite{Hurst1996Measures}, the frequent occurrence of accidents due to safety management failures highlights the need for audit tools. These tools can help evaluate an organization's safety climate and complement the assessment of safety management methods. By conducting audits of both safety climate and safety management practices, organizations can identify areas for improvement and enhance overall safety measures.\\
Indeed, there are similarities between good safety and health management practices and other management approaches such as quality management, environmental protection, and business excellence. Various management techniques can be applied across different areas, including safety and health management, quality management, environmental protection, and business excellence. Commercially successful companies often excel in occupational health and safety because they apply efficient business skills to safety and health management, just as they do for other aspects of their operations. This includes implementing effective risk assessment and mitigation strategies, establishing clear policies and procedures, providing adequate training and resources, promoting a culture of safety, and continuously improving safety performance through monitoring and evaluation. By integrating safety and health management into their overall business practices, these companies prioritize the well-being of their employees and create a safer working environment.

%% file: algorithm.tex

\section{A Comprehensive Overview of Brazil's Diverse Industries, Workforce, and Socioeconomic Landscape}

Brazil is indeed the world's fifth largest nation, covering an area of 8.5 million km2. It is also the sixth most populous country, with approximately 150 million inhabitants \cite{site}. The Brazilian economy is diverse and includes well-developed sectors such as mining, agriculture, manufacturing, and services. Mining plays a significant role in Brazil's economy, with rich mineral deposits supporting an extensive mining industry. Various resources are mined in different regions of the country. For example, there are major deposits of iron ore, bauxite, hematite, oil, and gas in many areas. According to De Medeiros Delgado et al.\cite{De1994Geo}, the Amazon region is known for its aluminum and manganese mining, while Bahia has chrome, magnesium, and quartz deposits. Copper and lead can be found in Bahia and Rio Grande do Sul, asbestos in Goias, and nickel in Goias and Minas Gerais. It's worth noting that Brazil is self-sufficient in tin, zinc, and tungsten. Additionally, the Amazon region is known for its gold deposits, which are mined by approximately 600,000 gold seekers, also known as ganimpeiros, who often work outside the formal sector. Overall, the mining industry in Brazil employs about 4 million people nationwide.\\
Brazil's manufacturing sector is indeed a rapidly growing and vibrant part of the economy. It encompasses a wide range of products, including steel, chemicals, petrochemicals, textiles, automobiles, pulp and paper, aircraft, electronics, and footwear. The petrochemical industry in Brazil is the largest in Latin America. The manufacturing sector has become a significant employer, currently employing about 25\% of the workforce and showing signs of continued expansion. Civil construction also employs a large number of workers. On the other hand, the service sector is the fastest-growing part of Brazil's economy and now employs over one-third of the workforce.\\
 In terms of workforce participation, approximately 45\% of Brazil's population, which is nearly 70 million people, is considered economically active, with around 60 million people being employed. Unfortunately, child labour remains a concern, with about 20\% of children aged 10 to 14 being part of the workforce (Population by situation in the labor market in Brazil \cite{Population}). Regarding gender distribution, women constitute about 35\% of Brazil's workforce. Female workers tend to have completed more schooling than their male counterparts. However, women are relatively more concentrated in the service sector and less concentrated in manufacturing, mining, agriculture, and construction compared to men.\\

 \subsection{Occupational Safety and Health Practice in Brazil}

Under the legal framework for occupational health regulation in Brazil, there are various components and entities involved. The federal constitution, issued in 1988, plays a significant role in this framework \cite{Menicucci2019Brazilian}. The Ministry of Labour has traditionally been responsible for most occupational health functions, but with the reorganisation of the health sector in 1988, some responsibilities were transferred to the Unified Health System (SUS). According to the new constitution, SUS has the authority to protect worker health, as stated in Article 205, Sections II and VII. However, the constitution does not provide specific guidelines on how this protection should be implemented. Additionally, no specific legislation defines the process for setting occupational health standards. The constitution does grant the federal government the power to conduct workplace inspections, as mentioned in Article 21, Section XXIV. The exact manner in which SUS will carry out regulation and workplace inspections under the new arrangement is yet to be determined. It is important to note that this information is based on the provided context and may not encompass all aspects of the legal framework for occupational health regulation in Brazil.\\
Currently, the Ministry of Labour in Brazil is responsible for occupational health regulation following the Consolidated Labour Law of 1943 \cite{Labor}. This law serves as the foundation for occupational health regulations in the country. In addition to federal regulations, state and municipal governments also have the authority to issue their own laws and regulations about occupational health. The Ministry of Labour has traditionally relied on threshold limit values (TLVs) established by the American Conference of Governmental Industrial Hygienists (ACGIH) as the basis for its regulations. These TLVs guide exposure limits to various substances in the workplace. It is important to note that this information is based on the provided context and may not encompass all aspects of occupational health regulation in Brazil.

%% file: numerical.tex
\section{Analytical Perspectives on Industrial Accidents }

The database contains records of 425 accidents from 12 different plants located in Brazil. By studying this data, organizations can gain insights into the occurrence of accidents, identify trends, and potentially uncover underlying causes. This analysis can help organizations identify areas for improvement in their safety and health management practices. By addressing these areas, organizations can enhance their safety measures, mitigate risks, and create a safer working environment for their employees. Organizations need to utilize this data effectively, employing appropriate statistical analysis and data visualization techniques to extract meaningful insights. This can aid in making informed decisions and implementing targeted improvements to prevent accidents and promote a culture of safety.\\ 
\begin{figure}[h!]
			\centering	
 \frame{\includegraphics[height=60mm]{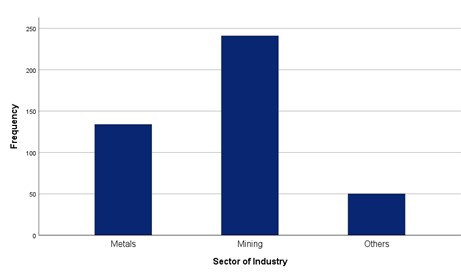}}
	\caption{The sector of industry and the frequency of accidents.}
	\label{fig:1}
	\end{figure}According to the depiction in Figure \ref{fig:1}, the mining industry accounted for 241 reported accidents, whereas the metals industry reported 134 accidents. Remarkably, a mere 50 accidents were reported across various other industries. \\
 There can be several reasons for this discrepancy: 
\begin{enumerate}
    \item Inherent risks: The mining industry is known to have inherent risks due to the nature of its operations, such as working in underground or hazardous environments, handling heavy machinery, or dealing with explosives. These factors can contribute to a higher likelihood of accidents compared to other industries.
    \item  Safety regulations: The mining industry often operates under strict safety regulations and protocols due to the potential for serious accidents and injuries. The reporting and documentation of accidents may be more comprehensive in this industry, leading to a higher number of reported incidents compared to other industries with less stringent regulations.
    \item Workforce size: The mining industry and the metals industry may have larger workforces compared to other industries, resulting in a higher number of reported accidents simply due to the larger number of workers.
    \item Industry-specific hazards: Each industry has its own unique set of hazards and risks. The specific hazards associated with mining and metals industries, such as cave-ins, equipment malfunctions, or exposure to toxic substances, may contribute to a higher number of accidents compared to other industries.

\end{enumerate}
\begin{figure}[h!]
			\centering	
 \frame{\includegraphics[height=70mm]{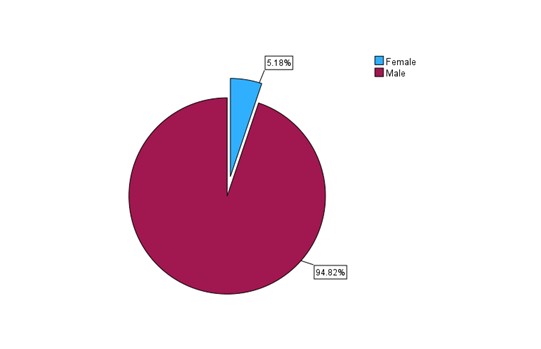}}
	\caption{The analysis of major industrial accidents by gender.}
	\label{fig:2}
	\end{figure}In Figure \ref{fig:2}, we demonstrate that accidents within the industry predominantly involve males, accounting for 94.82\% of reported incidents, while females make up only 5.18\%. This article aims to investigate and compare the involvement of males and females in these accidents, to understand the underlying causes, and to develop strategies to mitigate risks for both genders. \vspace{3cm}\\
Possible factors that contribute to this observation are:
\begin{enumerate}
    \item Occupational distribution: Men and women often have different job roles and occupational distributions within industries. Some industries, such as construction or mining, have a higher proportion of male workers, which may expose them to higher-risk tasks and environments.
    \item Physical differences: Men and women may have different physical capabilities and strengths, which can influence the types of tasks they perform. Certain physically demanding jobs may have a higher risk of accidents, and if these jobs are predominantly performed by men, it can contribute to the observed difference.
    \item Cultural and societal factors: Societal norms and expectations can influence the career choices and preferences of men and women. Men may be more likely to pursue jobs in industries with higher accident rates, while women may gravitate towards other sectors. This can contribute to the observed difference in accident rates.
    \item Safety culture and training: Workplace safety culture and training programs may not always address the specific needs and challenges faced by both men and women equally. This can result in a higher risk of accidents for men if safety measures are not effectively implemented or followed.

\end{enumerate}
It is important to emphasize that these factors are observations derived from analyzing databases gathered from mining, metal, and other industries, and may not have universal applicability. Efforts to promote gender equality in the workplace, improve safety measures, and provide comprehensive training can help reduce the disparity in accident rates between men and women in industries.
\begin{figure}[h!]
			\centering	
     \includegraphics[height=70mm]{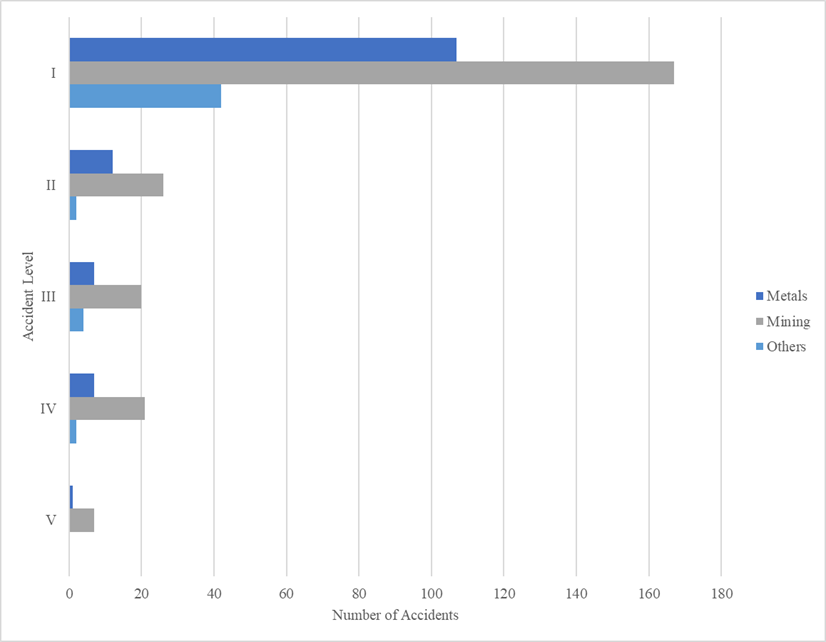}
	\caption{Analysis of the number of accidents by accident level.}
	\label{fig:3}
	\end{figure}Accidents in various industries are classified on a severity scale ranging from Category I to Category VI. Category I denotes milder accidents, while Category VI signifies extremely severe incidents. As depicted in Figure \ref{fig:3} in the metal industry, 107 accidents were recorded at the Category I level, along with 12 at the Category II level, 7 at the Category III level, 7 at the Category IV level, and 1 at the Category V level. Additionally, 167 accidents occurred in the mining industry at the Category I level, with 26 at Category II, 20 at Category III, 21 at Category IV, and 7 at Category V. Lastly, accidents in other industries accounted for 42 at Category I, 2 at Category II, 4 at Category III, and no accidents at Category IV or V. In total, there were 134 accidents in the metal industry, 241 in the mining industry, and 50 in other industries. This amounts to a total of 316 accidents at the Category I level, 40 at Category II, 31 at Category III, 30 at Category IV, and 8 at Category V.\vspace{0.3cm}\\ 
The mining industry is known to have a higher occurrence of dangerous accidents compared to other industries. Several reasons contribute to this:
\begin{enumerate}
    \item Hazardous working conditions: Mining operations often involve working in challenging environments such as underground mines, open pits, or remote locations. These conditions can expose workers to various hazards, including cave-ins, explosions, toxic gases, and the risk of being trapped or injured by heavy machinery.
     \item Complex machinery and equipment: The mining industry relies heavily on complex machinery and equipment, such as drills, crushers, and loaders. Operating and maintaining these machines require specialized skills and knowledge. Any malfunction or improper use of equipment can lead to severe accidents.
    \item Exposure to harmful substances: Miners may be exposed to harmful substances like dust, gases, and chemicals, depending on the type of mining being conducted. Prolonged exposure to these substances can lead to respiratory problems, occupational diseases, and other health issues.
    \item Geotechnical risks: Mining operations involve working with the earth's natural resources, which can pose geotechnical risks such as rockfalls, landslides, and collapses. These risks can be unpredictable and pose a significant danger to workers.

\end{enumerate}
Regarding the comparison between the mining and metal industries, it is worth noting that the mining industry involves extracting raw materials, while the metal industry involves processing and refining those materials. The metal industry may have a higher occurrence of accidents compared to other industries due to factors such as:
\begin{enumerate}
    \item High-temperature processes: Metal industries often involve high-temperature operations like smelting, casting, and welding. These processes can expose workers to burn hazards, fire risks, and the potential release of toxic fumes.
    \item Heavy machinery and equipment: Similar to the mining industry, the metal industry utilizes heavy machinery and equipment, which can pose risks if not operated or maintained properly.
    \item Handling of hazardous materials: Metal industries may involve working with hazardous materials such as acids, solvents, and heavy metals. Improper handling or exposure to these substances can lead to chemical burns, poisoning, or long-term health effects.

\end{enumerate}
Both the mining and metal industries must prioritize safety measures, provide comprehensive training, enforce strict regulations, and regularly assess and mitigate risks to ensure the well-being of workers and minimize accidents.\vspace{0.3cm}\\
\begin{figure}[h!]
			\centering	
 \frame{\includegraphics[height=70mm]{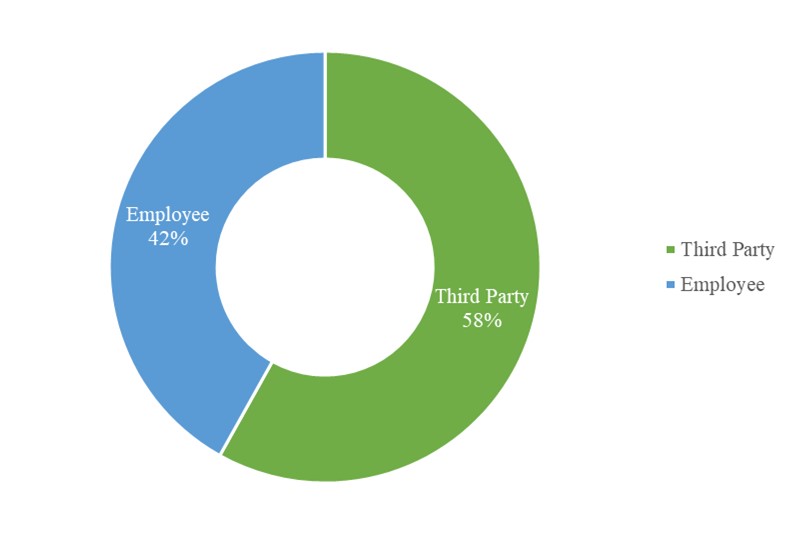}}
	\caption{Analysis of injured individuals by employment type.}
	\label{fig:4}
	\end{figure}Within this database, it was discovered that a significant number of accidents have affected both employees and third-party workers. Figure \ref{fig:4} stated that 178 employees and 257 third-party workers have experienced accidents in the database used to analyze accidents in mining, metals, and other industries.\\
Possible reasons for a higher occurrence of accidents among third-party workers compared to direct employees include:
\begin{enumerate}
    \item Lack of familiarity: Third-party workers may be less familiar with the specific work environment, equipment, and safety protocols compared to direct employees. This lack of familiarity can increase the risk of accidents.
    \item Limited training: Third-party workers may receive limited or insufficient training on safety procedures and protocols compared to direct employees. This can result in a higher likelihood of accidents due to a lack of knowledge or awareness.
    \item Time pressure: Third-party workers may face time constraints or deadlines imposed by their employers, which can lead to rushing or taking shortcuts. This can compromise safety measures and increase the risk of accidents.
    \item Communication challenges: Communication between third-party workers and the host company's employees may not always be as effective or clear. Misunderstandings or miscommunication regarding safety instructions or procedures can contribute to accidents.
    \item Lack of supervision: Third-party workers may have less direct supervision compared to direct employees. This reduced oversight can result in a lower adherence to safety protocols and an increased risk of accidents.

\end{enumerate}

%% file: conclusion.tex
\section{Conclusion}
A staggering number of industrial accidents in Brazil, particularly within the mining, metal, and other related industries. A total of 425 cases were observed, necessitating a detailed analysis to understand the root causes and potential preventive measures. This article aims to provide an in-depth examination of these accidents, shedding light on the key factors, the consequences, and the lessons learned.
Ensuring safety within an organization is of paramount importance in today's fast-paced and ever-evolving industrial landscape. Enterprises across all sectors must continuously strive to enhance their safety management practices and mitigate potential risks. This can be achieved through various means, including the activation of a safety culture within the organization, strengthening the determination of business owners, and providing education and training to workers.\\
It is evident that managing safety operating procedures and diligently addressing facility maintenance are critical to fostering a safe work environment. By placing a greater emphasis on these areas, organizations can enhance their preventive measures, reduce accidents, and safeguard the well-being of their workforce. Regularly reviewing and modifying safety procedures, along with providing comprehensive worker education, are paramount in ensuring the ongoing effectiveness of safety protocols. Additionally, daily inspections and effective equipment maintenance play pivotal roles in preventing accidents and creating a proactive safety culture within organizations.\\
The safety report submission process is a crucial aspect of preventing and managing chemical accidents. By maintaining and expanding this process to meet real-time demands and focusing on substances with high risk, we can significantly reduce the occurrence of accidents. Additionally, establishing a newly introduced system through the revision of the Occupational Safety and Health Act, along with efforts to establish a national safety culture, will further enhance overall safety. Let us prioritize safety and work towards creating a safer future for all.\\
Industrial safety measures have been successful in preventing major accidents caused by hazardous substance leakage, fires, and explosions. However, the persistent potential for accidents resulting from equipment defects or inadequate safety measures necessitates continuous vigilance. By adopting comprehensive preventive measures, conducting regular safety evaluations, and investing in ongoing training, industries can further ensure the safety and well-being of their workers and the broader community.\\
In modern industrial sites, safety is of utmost importance. Accidents not only cause harm to individuals but also lead to financial losses and damage a company's reputation. It is crucial to prioritize safety measures to ensure the well-being of employees and the smooth operation of industrial sites. The integration of auto-electronic systems has revolutionized safety practices, providing real-time monitoring and proactive action in hazardous situations. These systems utilize advanced technologies to enhance safety measures and minimize risks. implementing an auto-electronic safety management system offers numerous benefits to organizations. From improved safety measures and real-time monitoring to data-driven decision-making and cost savings, this advanced system revolutionizes safety practices. By embracing innovation and technology, companies can create a safer and more secure environment for their employees and stakeholders. With the advancement of auto-electronic safety management systems, the goal of zero accidents in industrial sites is within reach. By prioritizing safety and investing in proactive measures, companies can create a secure work environment that safeguards the well-being of employees and facilitates uninterrupted operations. Embracing these technologies and adopting a safety-first approach is crucial for future success in the industrial sector.